\documentclass[manuscript]{aastex}
\usepackage{topcapt}
\shorttitle{SN 2007od: A Type IIP SN with Circumstellar Interaction}
\shortauthors{Andrews et al.}
\usepackage{booktabs}
\usepackage{natbib}

\begin{document}

\title{SN 2007od: A Type IIP SN with Circumstellar Interaction}

\author{J.E. Andrews\altaffilmark{1},  J.S. Gallagher\altaffilmark{1}, Geoffrey C. Clayton\altaffilmark{1}, B.E.K. Sugerman\altaffilmark{2}, J.P. Chatelain\altaffilmark{1}, J. Clem\altaffilmark{1}, D.L. Welch\altaffilmark{3}, M.J. Barlow\altaffilmark{4}, B. Ercolano\altaffilmark{4,5}, J. Fabbri \altaffilmark{4}, R. Wesson \altaffilmark{4}, and M. Meixner\altaffilmark{6}}

\altaffiltext{1}{Department of Physics and Astronomy, Louisiana State University, 202 Nicholson Hall, Baton Rouge, LA 70803; jandrews@phys.lsu.edu, jgallagher@phys.lsu.edu, gclayton@fenway.phys.lsu.edu, jchate6@tigers.lsu.edu, jclem@phys.lsu.edu}
\altaffiltext{2}{Department of Physics and Astronomy, Goucher College, 1021 Dulaney Valley Rd., Baltimore, MD 21204; ben.sugerman@goucher.edu}
\altaffiltext{3}{Department of Physics and Astronomy, McMaster University, Hamilton, Ontario, L8S 4M1 Canada; welch@physics.mcmaster.ca}
\altaffiltext{4}{Department of Physics and Astronomy, University College London, Gower Street, London WC1E 6BT, UK; mjb@star.ucl.ac.uk, jfabbri@star.ucl.ac.uk, rwesson@star.ucl.ac.uk}
\altaffiltext{5}{Institute of Astronomy, University of Cambridge, Madingley Road,Cambridge, CB3 0HA; be@ast.cam.ac.uk}
\altaffiltext{6}{Space Telescope Science Institute, 3700 San Martin Drive, Baltimore, MD 21218; meixner@stsci.edu}

\begin{abstract}
SN 2007od exhibits characteristics that have rarely been seen in a Type IIP supernova (SN).  Optical $V$-band photometry reveals a very steep brightness decline between the plateau and nebular phases of $\sim$4.5 mag, likely due to SN 2007od containing a low mass of $^{56}$Ni. The optical spectra show an evolution from normal Type IIP with broad H$\alpha$ emission, to a complex, four component H$\alpha$ emission profile exhibiting asymmetries caused by dust extinction after day 232. This is similar to the spectral evolution of the Type IIn SN 1998S, although no early-time narrow ($\sim$200 km s$^{-1}$) H$\alpha$ component was present in SN 2007od. In both SNe, the intermediate-width H$\alpha$ emission components are thought to arise in the interaction between the ejecta and its circumstellar medium (CSM). SN 2007od also shows a mid-infrared excess due to new dust. The evolution of the H$\alpha$ profile and the presence of the mid-IR excess provide strong evidence that SN 2007od formed new dust before day 232. Late-time observations reveal a flattening of the visible lightcurve. This flattening is a strong indication of the presence of a light echo, which likely accounts for much of the broad, underlying H$\alpha$ component seen at late-times. We believe the multi-peaked H$\alpha$ emission is consistent with the interaction of the ejecta with a circumstellar ring or torus (for the inner components at $\pm$1500 km s$^{-1}$) , and a single blob or cloud of circumstellar material out of the plane of the CSM ring (for the outer component at -5000 km s$^{-1}$). The most probable location for the formation of new dust is in the cool dense shell created by the interaction between the expanding ejecta and its CSM. Monte Carlo radiative transfer modeling of the dust emission from SN 2007od implies  that up to $\sim$4 x 10$^{-4}M_{\sun}$ of new dust has formed. This is similar to the amounts of dust formed in other core-collapse supernovae such as SNe 1999em, 2004et, and 2006jc.

\end{abstract}


\section{Introduction}

The discovery of massive amounts of dust in luminous infrared (IR) high-z galaxies, seen less than 1 Gyr after
the Big Bang \citep{2003A&A...406L..55B}, challenges our understanding of major dust contributors in the
Universe. After only 1 Gyr, there has not been time for low-mass stars to form and evolve to the asymptotic giant branch, which accounts for about half of the dust production in the Galaxy at the current epoch \citep{1998ApJ...501..643D}.
Therefore, it has been suggested that core-collapse supernovae (CCSNe) may be significant contributors to the dust production in high-z galaxies. Surprisingly, dust mass estimates derived from the observed SEDs (spectral energy distributions) of recent SNe have been much lower than necessary to account for  the dust observed in early-time galaxies.  The sample of measured SNe is small but it appears that a typical Type II SN is producing only 10$^{-4}-10^{-5}M_{\sun}$ of dust, well below the 1 M$_{\sun}$ needed to account for the large amounts of dust seen in the early Universe \citep{1989ApJ...344..325K, 2001MNRAS.325..726T}. SN 2004et has been found to have formed only a few 10$^{-4}M_{\sun}$ of dust  \citep{2009ApJ...704..306K}. \citet{2006Sci...313..196S} showed that larger masses (10$^{-2}M_{\sun}$) of dust can be derived in the case of SN 2003gd if the dust is considered to be clumpy and mostly silicate, but \citet{2007ApJ...665..608M} questioned these findings and derive smaller dust masses (10$^{-4}M_{\sun}$) for SN 2003gd. 
Finally, \citet{2007MNRAS.375..753E}, using similar methods to \citet{2006Sci...313..196S}, did not find that large amounts of dust had condensed in the Si-poor envelope of SN 1987A.

There are three distinct signatures of dust formation in CCSNe. These include a drop in visual brightness due to increased dust extinction accompanied by an infrared (IR) excess, and the appearance of asymmetric emission line profiles caused by the greater attenuation by dust in the ejecta of redshifted emission than blueshifted emission.  These signatures usually appear a few hundred days after the explosion and only SN 1987A \citep{1989LNP...350..164L,1993ApJS...88..477W}, SN 2003gd \citep{2006Sci...313..196S, 2007ApJ...665..608M}, and SN 2004et \citep{2006MNRAS.372.1315S,2009ApJ...704..306K} have shown all three. 

A few CCSNe have shown evidence for early dust formation. SN 1998S formed dust sometime between days 140-268 \citep{2000ApJ...536..239L}, while SN 2006jc showed dust formation signatures between days 50-75 \citep{2008MNRAS.389..141M,2008ApJ...680..568S}.  The SNe which show signs of early dust formation are all of type IIn, where the ejecta interacts with the circumstellar medium (CSM) of the SN. The dust may form in a cool dense shell (CDS) between the forward and reverse shocks \citep{1994ApJ...420..268C,2004MNRAS.352..457P}. The formation of dust in the CDS allows for an additional avenue of dust formation not available in normal Type II SNe. The capability to form dust both within the SN ejecta and in the CDS suggests that these interacting systems have potential for forming dust in greater quantities than normal Type II SNe, although to date they have produced dust amounts comparable to normal Type II SNe.    In this paper, we present new spectroscopic and photometric data covering 800 days of the evolution of SN 2007od, an unusual Type IIP SN that shows evidence for CSM interaction and early-time dust formation.

\section{Observations}
\objectname{SN 2007od} is located in \objectname{UGC 12846} at a distance of 24.5 Mpc \citep{2007ATel.1259....1I}. It was discovered on 2007 November 2 with $R$ $\sim$14.4 mag \citep{2007CBET.1116....1M}, and two days later was confirmed spectroscopically as a normal Type II SN most closely resembling SN 1999em at 10 days past explosion \citep{2007CBET.1119....1B}.  Based on this, we assume an explosion date of 2007 October 25 (JD 2454398) throughout this paper. The new observations reported in this paper are summarized in Table 1. We have obtained visible spectroscopy and photometry spanning days 57-811 as well as mid-infrared photometry on days 300, 455 and 667 of SN 2007od.  

Optical imaging was obtained in Johnson-Cousins $BVI$  with the SMARTS consortium 1.3m telescope at Cerro Tololo Inter-American Observatory (CTIO), Chile.  All images were pipeline reduced, shifted, and stacked. Imaging and spectra were also obtained with Gemini/GMOS North and South (GS-2008A-DD-2, GS-2008B-Q-45, GN-2008B-Q-245, GS-2009B-Q-1). The g$^{\prime}$r$^{\prime}$i$^{\prime}$ images were reduced and stacked using the IRAF \textit{gemini} package.  The instrumental g$^{\prime}$r$^{\prime}$i$^{\prime}$  magnitudes were transformed to standard Johnson-Cousins $VRI$ \citep{2007ApJ...669..525W,1996AJ....111.1748F}.  For each night the transformation involved a least-squares fit with a floating zero point.  Late time images were obtained with the Wide Field Camera (WFC) on HST/ACS using the F435W, F606W, and F814W filters. These images were pipeline reduced, including drizzling and cosmic ray removal, and transformations to the Johnson-Cousins $BVI$ system were accomplished using \citet{2005PASP..117.1049S}. The large bandpass of the F606W filter includes the 6563 \AA\ H$\alpha$ emission, which is the main source of the late-time luminosity.  We believe this causes the transformed Johnson $V$ magnitude from HST to appear brighter than the $V$ magnitude obtained from the Gemini transformations. In the Gemini filter set, r$^{\prime}$ contains the H$\alpha$ flux, and is a very small component to the Johnson $V$ transformation but a large component to the $R$ transformation. Therefore the $R$ magnitude from Gemini would contain the H$\alpha$ emission whereas H$\alpha$ would be present in the $V$ magnitude from HST. 

A $BVRI$ photometric sequence of tertiary standard stars (shown in Figure 1) was derived for the SN 2007od field. A table of magnitudes for these standards is located in Appendix A. The $VRI$ lightcurves of SN 2007od are shown in Figure 2. Uncertainties for the Gemini images were calculated by adding in quadrature the transformation uncertainty quoted in \citet{2007ApJ...669..525W}, uncertainty from photon statistics, and the zero point deviation of the standard stars for each epoch. The HST uncertainties consist of the transformation uncertainty from \citet{2005PASP..117.1049S} and the uncertainty from photon statistics added in quadrature. 
We used the narrow Na I D absorption seen in the day 10 spectrum of SN 2007od (described below) to estimate the foreground extinction. The equivalent width of the blended Na I D lines is 0.50 $\pm$0.05\AA. This corresponds to A$_{v}\sim$0.39 $\pm$0.04 \citep{1990A&A...237...79B}. Since the value is small, the photometry presented in Figure 2 and Table 2 have not been foreground corrected.

For the first three Gemini/GMOS South epochs, three integrations of 900s were obtained in longslit mode using grating B600 and a slit width of 0$\farcs$75.  For the last two epochs on days 672 and 699, six integrations of 1800s each were obtained. Central wavelengths of 5950, 5970, and 5990 \AA\ for the first three GMOS-S epochs were chosen to prevent important spectral features from falling on chip gaps, and central wavelengths of 6350, 6370, 6390 \AA\ were chosen for the last two epochs to also include the  [Ca II] $\lambda\lambda$7291,7324 emission. A 2x2 binning in the low gain setting was used. The one GMOS-N epoch used the R400 grating with central wavelengths of 5460, 5480, and 5500 \AA. All other settings were the same as in the early GMOS-S epochs. Spectra were reduced using the IRAF \textit{gemini} package.  The sky subtraction regions were determined by visual inspection, and included 30 pixels on either side of the SN. The spectra were extracted using 20 rows centered on the SN. The spectra from each individual night were averaged and have been corrected for the radial velocity of  \objectname{UGC 12846} (1734 km s$^{-1}$). The spectra obtained on days 672 and 699 have been combined to increase the S/N. The spectra on days 10 and 50 were taken with the FAST spectrograph operating on the 1.5-m Tillinghast telescope at the F. L. Whipple Observatory (St\'{e}phane Blondin, personal communication). The spectra are shown in Figures 3 and 4.

We obtained three epochs of Spitzer/IRAC imaging and one epoch each with IRS PUI Blue and MIPS. See Table 1 and Table 3. The Spitzer IRAC (3.6, 4.5, 5.8, and 8.0 $\mu$m), IRS PUI Blue (16 $\mu$m), and MIPS 24 $\mu$m images were mosaicked and resampled using standard MOPEX procedures to improve photometric quality. Aperture photometry was performed utilizing the prescribed aperture corrections obtainable from the Spitzer Science Center website$^{1}${\footnotetext[1]{http://ssc.spitzer.caltech.edu/obs/}.  Figure 5 presents the SED of SN 2007od at three epochs, Spitzer (day 300)/Gemini (day 309), Spitzer (day 455-471)/Gemini (day 458), and warm Spitzer (day 667)/Gemini (day 672). The data have not been corrected for foreground extinction. Statistical uncertainties presented in the figure represent 1$\sigma$ errors. It should be noted that the SN was not detected in the MIPS and IRS/PUI images obtained on days 471 and 455, respectively. Consequently, these data are upper limits. Table 3 lists the measured mid-IR fluxes of SN 2007od.

\section{Discussion}

\subsection{Lightcurve Evolution}
A typical Type IIP SN lightcurve shows a plateau for the first $\sim$60-100 days, followed by a sharp drop in luminosity of $\sim$1.5-3 mag as the SN transitions into the nebular phase, after which the lightcurve exhibits an exponential decline powered by $^{56}$Co decay \citep{1986ARA&A..24..205W}.
The lightcurve of SN 2007od (Figure 2) shows a normal plateau phase followed by an exceptionally large drop in brightness as it is entering the nebular phase. The last photometry before the gap was obtained on day 82 (Chornock 2009, personal communication). If we assume the end of the plateau phase took place shortly after that time and we
extrapolate the slope of the $^{56}$Co decay  backwards into the gap (days 82-232) when no photometry is available, we find that the $V$ brightness fades a minimum of 4.5 mag over the typical twenty day drop period  \citep{2003A&A...404.1077E}.
\citet{1986ARA&A..24..205W}, and more recently \citep{2009ApJ...703.2205K}, have suggested that if only a small amount of  $^{56}$Ni is synthesized in the SN explosion the lightcurve will show a steep drop between the plateau and nebular phases due to a lack of sufficient thermal energy. 
Generally, such a steep drop is only seen in low-luminosity (M$_{v}\sim$-14) SNe such as 1999br, 1999eu and 2005cs, that have very low inferred $^{56}$Ni masses (2 -- 8 x 10$^{-3}$M$_{\sun}$)  \citep{2004MNRAS.347...74P}. SN 1994W, which dropped $\sim$3.5 mag from days 110 to 122 \citep{1998ApJ...493..933S} is the only other luminous (M$_{v}\sim$-18) Type IIP besides SN 2007od to show such a steep drop in brightness. Its $^{56}$Ni mass was estimated to be $\sim$2.6 x 10$^{-3}$M$_{\sun}$. \citet{1998ApJ...493..933S} suggested that after its 3.5 mag drop from the plateau, SN 1994W declined much faster than the $^{56}$Co decay once in the nebular
phase, possibly due to dust formation. 

After day 458 the lightcurve of SN 2007od shows a deviation from  $^{56}$Co decay in the form of a flattening of the lightcurve between days 659 and 811. By day 811, assuming a constant brightness has been maintained from day 672, the difference in an extrapolated $^{56}$Co decay is then $\sim$2.2 magnitudes in $V$. At late times, the Gemini and HST $V$ magnitudes differ significantly due to differences in the filter bandpasses as discussed in Section 2. However, the constancy of the three epochs of HST/ACS $V$ and $I$ photometry from day 659 to 811 show that the flattening of the lightcurve is real. It is likely being caused by a light pulse from the SN reflecting off of surrounding material and creating a light echo. Given the distance of the SN (24.5 Mpc), platescale of the WFC detector (0\farcs05~pix$^{-1}$) and FWHM of the stellar PSF (1.9 pix), a source could have a spatial extent on the plane of the sky of up to 0.12\arcsec or 46 lt-yr and still be unresolved on the detector.  In particular, we find through experimentation that a source with the flux of this SN and with a FWHM up to 1.2 times that of the PSF will not leave detectable residuals when subtracted away, therefore we allow the physical size of the emitting region to be up to 0.12\arcsec.  Using the light echo equation \citep{1939AnAp....2..271C,2003AJ....126.1939S}, dust could be up to 475 lt-yr in front of the SN on day 811 and still remain unresolved in the HST imaging.  Thus, the echoes could be produced from either circumstellar or interstellar dust.  Given that the foreground extinction to the SN is low ($A_V \sim 0.4$), any dust producing a light echo should be optically thin. Typically, the contribution from a light echo becomes significant when a SN has faded $\sim$8 mag from maximum \citep{2006MNRAS.369.1949P}. SN 2007od was more than 10 mag below maximum light by day 672  when the contribution of the echo to the lightcurve became significant. A qualitative comparison of the lightcurve and evolution at late times of SN 2007od to models by \citet{2005MNRAS.357.1161P} suggest that the echo is more likely due to an interstellar sheet than circumstellar dust.  We discuss possible contributions of an echo to the spectral evolution of SN 2007od  in Section 3.2.  

As seen in Figure 2, the evolution of the $V-I$ colors is normal for a dust forming Type IIP SN. We have only included the $V-I$ colors from the Gemini observations due to the bandpass considerations of HST described in Section 2. At the time of dust formation, previous photometry of SNe 1999em and 2004et showed only small ($<$0.1 mag) changes in their $V-I$ colors.  The $V-I$ color evolution of SN 2007od before and after the gap in observations (days 82-232) when we think dust formation began, is consistent with these other SNe. The lack of change in the $V-I$ color evolution indicates that, in general, the color evolution is not sensitive to dust formation. This is likely caused by either a very small amount of dust being formed, or larger amounts forming in clumps. If the dust is clumpy, then the color evolution would not reflect the optical depth of dust present and the colors would not be as reddened as they would be if the dust were uniformly distributed e.g., \citet{1996ApJ...463..681W}.

\subsection{Spectral Evolution}

The spectral evolution of SN 2007od shows behavior not previously seen in other SNe. Figure 3 shows the full optical spectrum of SN 2007od at 10, 232, and 458 days. Early-time spectra showed SN 2007od to be a normal Type II SN with a blue continuum, broad H$\alpha$ emission with a P-Cygni profile, and an expansion velocity of $\sim$10000 km s$^{-1}$  \citep[Chornock 2009, personal communication]{2007CBET.1119....1B}. Other notable spectral features at early times include broad H$\beta$ emission, weak Na I D $\lambda\lambda$5890,5896 absorption, and [O I] $\lambda\lambda$6300,6363 emission that was weaker than that normally seen in Type II SNe. There was no sign of either a narrow (FWHM  $\sim$ 200 km s$^{-1}$) H$\alpha$ component nor X-ray emission \citep{2007CBET.1119....1B,2007ATel.1259....1I}, typically seen in a Type IIn SN.   If present, the X-ray luminosity of SN 2007od was $<$ 1.0 $\times$ 10$^{40}$ erg s$^{-1}$, significantly smaller than luminosities measured in Type IIn SNe such as 2006jd and 2007pk \citep{2007ATel.1290....1I,2007ATel.1284....1I}.

By the time of the first Gemini/GMOS epoch (day 232) the spectrum of SN 2007od showed evidence for strong interaction with its CSM, an event usually associated with Type IIn SNe.  The H$\alpha$ emission profile had changed dramatically from the early-time spectra, with the single broad component evolving into at least three intermediate width (FWHM $\sim$ 1700 km s$^{-1}$) emission components on top of a fainter, broader emission component (Figure 4).  Two of these intermediate width components are arranged symmetrically about zero velocity at  $\pm$1500  km s$^{-1}$ and the third, which becomes more prominent at later epochs, lies at -5000 km s$^{-1}$. We should note that the narrow emission at zero velocity is HII emission from the host galaxy not associated with the SN.  Underlying the three intermediate width components, the broad H$\alpha$ emission present in the early-time spectra can still be seen, and it is likely that it is responsible for the broad red wing of the H$\alpha$ emission profile (see Figure 4). Broad, blended Fe II emission between 5100-5400 \AA\, similar to that seen in SN 1998S \citep{2004MNRAS.352..457P} appeared by day 232, and [Ca II] $\lambda\lambda$7291,7324 emission became prominent by day 458 (Figure 3).

A similar multi-component structure was seen in the H$\alpha$ profile of SN 1998S.  Unlike SN 2007od, SN 1998S was classified as a Type IIn, and showed both broad and narrow H$\alpha$ emission components at early-times due to the expanding ejecta and photoionized CSM material, respectively. The narrow component disappeared before day 108 \citep{2000ApJ...536..239L}, and sometime before day 250, a triple-peaked H$\alpha$ emission profile had developed \citep{2000AJ....119.2968G} (Figure 4).  The three intermediate width components persisted in SN 1998S from approximately day 250 to day 660, exhibiting a progressive fading of the red and central components with respect to the blue component; a phenomenon that was attributed to dust formation \citep{2000AJ....119.2968G}.  The timescale for the evolution of the SN 1998S spectrum is similar to that of SN 2007od, which showed a single broad H$\alpha$ emission component from days 10-114, followed by a period (114-232 days) when the SN was unobservable, and finally the emergence of  multiple-component emission by day 232.  In SN 2007od, the blueshifted component  at -1500 km s$^{-1}$ is much stronger than its corresponding component at +1500 km s$^{-1}$.  Like SN 1998S, the simplest explanation for this asymmetry is that newly formed dust is extinguishing the redshifted, far-side emission more than the blueshifted, near-side emission.  The ratios of the inner peak heights would require an optical depth of approximately 1.0 at H$\alpha$ for the asymmetry to be attributable to dust attenuation.  If the dust existed in the CSM before the SN explosion, then it would lie outside of the ejecta and all of the emission components would be equally reddened.  It is thought that the outer peaks in the H$\alpha$ profile of SN 1998S originated in the interaction of the ejecta with a circumstellar ring or torus \citep{2000AJ....119.2968G,2000ApJ...536..239L,2002ApJ...572..932P,2004MNRAS.352..457P,2005ApJ...622..991F}.  If we assume the same is true of the asymmetric inner peaks of SN 2007od, then there must be dust between the source of the near-side (blueshifted) and the far-side (redshifted) emission components in the torus (Figure 6). Moreover, if this dust is forming in the CDS, as was suggested by \citet{2004MNRAS.352..457P} for SN 1998S, then the emission would likely be coming from a radiative forward shock \citep{2004MNRAS.352..457P,2005ApJ...622..991F} as one would expect equal attenuation of both redshifted and blueshifted emission originating in the reverse shock (see Figure 6). It is also worthwhile to note that \citet{2009ApJ...704..306K} has recently alluded to CSM interaction producing a similar multi-component H$\alpha$ profile in SN 2004et, but occuring at least 400 days after it was seen in SNe 1998S and 2007od.

In SN 2007od,  the $\pm$1500  km s$^{-1}$ peaks appear sometime between day 114 and 232.  Given the maximum velocity of the ejecta of 10000 km s$^{-1}$, this constrains the distance of the inner radius of the CSM responsible for these profile peaks to $\sim$700-1300 AU.  Assuming a progenitor wind of 10 km s$^{-1}$, this CSM was formed during a mass loss event ending 300-600 years prior to the SN explosion.  With the emergence of the -5000 km s$^{-1}$ components between days 114 and 309, a similar analysis places this CSM at 700-1700 AU and the formation period between 300-800 years prior to explosion.

The appearance of the asymmetric components between days 114 and 232 implies that dust formation began much earlier than typically seen for non-interacting CCSNe.  These SNe (SN 1987A, 2003gd, 2004et) showed developing asymmetries between days 300 and 600 post explosion  due to dust condensation in the ejecta \citep{1989LNP...350..164L,2006Sci...313..196S,2006MNRAS.372.1315S}.  Furthermore, the [Ca II] $\lambda$7291 emission profile in the day 458 spectrum shows only the one broad component, with no intermediate-width components present (Figure 3). This implies that the complex emission seen in H$\alpha$ is arising in the hydrogen-rich CSM lost from the star before the explosion and not in the metal-enriched ejecta \citep{2000AJ....119.2968G}. These observations are more consistent with dust formation in the CDS as has been suggested in SN 1998S \citep{2004MNRAS.352..457P,2005ApJ...622..991F}.

The H$\alpha$ profile of SN 1998S also showed a third peak near zero velocity that has been attributed to the interaction of the ejecta with a spherical distribution of shocked CSM clouds \citep{2000AJ....119.2968G,2000ApJ...536..239L,2002ApJ...572..932P,2004MNRAS.352..457P,2005ApJ...622..991F}.  In the case of SN 2007od, Figure 4 also shows the emergence of a third peak at -5000  km s$^{-1}$ that was hidden by the broad ejecta emission on day 232.  This blue peak becomes increasingly more prominent as the broad underlying ejecta emission begins to fade over time.   By day 699, when we believe the light echo is contributing significantly to the flux, this feature is still prominent. Because a light echo depicts only the early-time emission, it is not possible to explain the -5000 km s$^{-1}$ peak with a light echo scenario. The nature of this third peak represents the starkest difference between SNe 2007od and 1998S.  As previously mentioned, in SN 1998S the peak not associated with the CSM torus was observed at zero velocity and attributed to a spherical distribution of shocked CSM clouds.  Although a similar solution can not be invoked to explain a peak centered at -5000 km s$^{-1}$ as seen in SN 2007od, it is possible that this peak comes from the emission of a single cloud or blob of shocked CSM.  If this scenario is correct, because the torus and the blob are seen at different velocities, the blob cannot be located in the plane of the torus. At the same time, the blob would be required to have a velocity component along our line of sight larger than that of the torus.  Therefore, if the torus is nearly edge on as viewed from Earth (Figure 6), as hypothesized in SN 1998S, the blob cannot be moving orthogonal to the plane of the torus as this would severely limit its velocity component along our line of sight.  On the other hand, if the inclination$^{2}${\footnotetext[2]{\textit{i} = 90$^{\circ}$ is edge-on} of the ring is less than 90$^{\circ}$, the blob could be moving orthogonally, but it would require a sufficiently thick torus to retain the dust extinction necessary to explain the asymmetric inner peaks. The scenario presented here to account for the observations is closely based on the scenario used to explain the observations of SN 1998S. This torus + blob scenario is introduced in order to illuminate the manner in which the multi-component H$\alpha$ emission may arise, but the existing data do not allow the detailed geometry of SN 2007od to be ascertained.

From days 232 to 699, the relative strength of the emission component at -5000 km s$^{-1}$ is increasing as the inner pair of emission components ($\pm$ 1500 km s$^{-1}$) slowly disappear. Again if we assume the SN 1998S scenario, then this is explained by the ejecta moving through, and completely destroying, the torus but
not the blob. The underlying broad component is also present at all epochs. The broad component is likely a combination of the residual ejecta emission and, given the flattening of our lightcurve $\sim$10 mag below maximum, a light echo.  The contribution to the total broad component by the light echo is expected to grow with time as the ejecta emission fades. Based on the lightcurve, on day 238 the light echo flux would be $\sim$ 20x fainter than the directly transmitted ejecta emission, and on day 348 the echo would still be $\sim$8x fainter. However, by day 672/699, the echo flux will likely be greater than the directly transmitted ejecta emission. Therefore the echo should not be a major contributor to the H$\alpha$ profile at early times, but would be a major contributor of the late-time profile. Figure 4 shows the fits to the day 232 and 348 H$\alpha$ profiles.  We have used Gaussians to approximate the intermediate width peaks and a scaled day 50 spectrum to approximate the residual ejecta emission.  We should note that the scaled day 50 spectrum is only an approximation to the day 232 and 358 ejecta emission.  Realistically, in addition to fading over time, the profile would likely lose its P Cygni absorption, and its overall width would likely decline. The flat-topped profile of the -5000 km s$^{-1}$ component is shown to be well fit by two intermediate width Gaussians possibly due to density enhancements within the CSM cloud or blob.  At later times the profile becomes too complex to fit.  On days 458 and 672/698, the -5000 km s$^{-1}$ peak begins to dominate and its two component nature becomes more clear.  Moreover, there is a continuing fading of the inner peaks to the point where they  disappear into the broad emission by day 672/698.  As previously mentioned, by this last epoch, the broad emission is a combination of the light echo spectrum and the weak residual ejecta emission.  However, the residual flux beyond 5000 km s$^{-1}$ is unlikely to originate from either of these two sources and remains unexplained. There is also an indication that the radial velocities of both sets of components get smaller with time. This is similar to the behavior of the blue component of SN 1998S which moved to smaller velocities from day 249 to day 658, and was explained as an interaction of the shock with more massive sections of CSM \citep{2004MNRAS.352..457P}.

\subsection{Radiative Transfer Modeling}

The SEDs of SN 2007od shown in Figure 5 all show strong IR excesses attributable to warm dust. The SED at 300 days can be fitted well by the sum of a blackbody with a temperature of 5100 K representing hot, optically-thick gas in the ejecta and a modified blackbody with a temperature of 580 K subject to a $\lambda$$^{-1}$ emissivity law representing the dust.  A similar fit to the day 455 SED uses a 5100 K blackbody and a 490 K modified blackbody, and for day 667 a 5100 K blackbody and a 600 K modified blackbody.  The fits are shown in Figure 5 and the relevant parameters are compiled in Table 4.  The IR emission could arise from pre-existing dust in the CSM, and/or new dust condensing in either or both of the CDS and the ejecta itself.  IR echoes from pre-existing CSM dust heated by the early-time luminosity are often seen in other SNe. The position of the dust in SN 2007od cannot be inferred from IR observations alone, but when combined with the evidence provided by the asymmetry in the H$\alpha$ emission components, it suggests that the IR emission is coming from dust forming in the ejecta or the CDS, and not from an IR echo.  The very early onset of dust formation (between 114-232 days) further suggests that the dust formation is occurring in the CDS since the ejecta at this time would still be too warm for dust condensation \citep{1991A&A...249..474K}.

In order to better understand the amount and composition of the dust forming within SN 2007od we used our 3D Monte Carlo radiative transfer (RT) code MOCASSIN \citep[and references therein]{2005MNRAS.362.1038E}.  For the purposes of our RT models we will assume the dust and source luminosity is spread throughout a spherical, expanding shell with an inner radius, R$_{in}$, and an outer radius radius, R$_{out}$.  The luminosity at each location is proportional to the local density. With this simple model, we can quantify the total amount of dust in the system as a whole (ejecta + CDS) and monitor its variation from day 300 to 667.  We have modeled two different dust distributions.  Following radiative transfer models of other Type II SNe \citep{2006Sci...313..196S,2007MNRAS.375..753E}, the ``smooth" model for SN 2007od has the dust uniformly distributed throughout the shell according to a r$^{-2}$ density profile, while the ``clumpy"  model reflects an inhomogeneous distribution of spherical clumps embedded within an interclump medium of density $\rho$.  The irradiating photons are only produced in the interclump medium, while the clumps are assumed to be dark (see \citet{2007MNRAS.375..753E} for further discussion of this issue). We have adopted a standard grain size distribution of $a^{-3.5}$ between 0.005 and 0.05 $\mu$m  \citep{2006Sci...313..196S, 2007ApJ...665..608M,2009ApJ...704..306K} and have tested two different concentrations, i.e. a dust model dominated by silicates (75\% Si; 25\% AC) and one dominated by amorphous carbon (75\% AC; 25\% Si).  For each of these models, we used the optical grain constants of \citet{1988ioch.rept...22H} and \citet{1984ApJ...285...89D}, respectively.

The main input parameters for the model are the ejecta temperature and luminosity, the inner and outer radii of the shell, and the mass of dust present.  Initial values of the luminosity, temperature, and approximate size of the dust shell are based on the results of our blackbody analysis above.  Tuning the dust mass, along with slight variations in the other parameters, allowed us to obtain a good fit to the optical and Spitzer/IRAC points for the first two epochs. We should point out that without the 5.8 and 8.0 $\mu$m data on day 667, we were unable to make reliable constraints on the dust mass at this epoch. The best fit to the data was obtained with a smooth dust distribution and is shown in Figure 5 and the corresponding input parameters are listed in Table 5. This table also lists the dust mass calculated by our clumpy model.

Although no attempts were made to fine tune the models to match the 16 and 24 $\mu$m flux upper limits on day $\sim$455, care was taken to ensure that our preferred model did not predict emitted fluxes that would have been detected given our experimental setup.  We found that we were unable to account for this relatively low flux at 16 and 24 $\mu$m with compositions dominated by silicate dust (see Figure 5).  All other parameters being equal, transitioning from amorphous carbon dust to silicate dust has the effect of broadening the IR peak as the SED will then be increased by strong silicate emission features at 10 and 18 $\mu$m.  These two emission bumps are clearly seen in the Si model in Figure 5.  The conclusion is that the silicate dominated model predicts too much flux beyond 10 $\mu$m at day 455 leading us to favor the amorphous-carbon dominated model with 75\% AC and 25\% Si.  Given this concentration, our best smooth model predicts a total of 1.7 x 10$^{-4}$ M$_{\sun}$ and 1.9 x 10$^{-4}$ M$_{\sun}$ of dust on days 300 and 455, respectively, with an increase in $\tau$$_{v}$ from 1.5 to 2.0 over the two epochs. This implies a total of 2 x 10$^{-5}$ M$_{\sun}$ of new dust forming between day 300 and 455 after explosion.  The values of $\tau$ at 6600 \AA\ predicted by our MOCASSIN fits were 1.3 and 1.8 for days 300 and 460, respectively. These opacities are adequate to explain the inner peak height ratios discussed in Section 3.2.  Although the constraints were less reliable on the dust mass at day 667, we were able to show that the available data is consistent with our previous epochs with a predicted dust mass of 1.8 x 10$^{-4}$ M$_{\sun}$ and $\tau$$_{v}$ = 2.75.  The best clumpy model gave a total dust mass of 2.5 x 10$^{-4}$ M$_{\sun}$ and 4.2 x 10$^{-4}$ M$_{\sun}$ on days 300 and 455/667, respectively, implying the formation of 1.7 x 10$^{-4}$ M$_{\sun}$ of new dust formed. In each case, the quantity of dust formed is several orders of magnitude lower than that required to contribute significantly to the dust budget of the early universe \citep{2003MNRAS.343..427M}.

\section{Summary}

SN 2007od is a very unusual Type IIP SN that shows strong evidence for early-time ($\lesssim$232 days) dust formation in the cool dense shell (CDS) created by the interaction between the ejecta and the circumstellar medium (CSM). Observations prior to day 82 for the optical lightcurve and prior to day 114 for the spectral evolution indicate SN 2007od was a normal Type IIP SN. Photometry and spectra obtained on day 232 revealed that a large post-plateau luminosity drop ($\sim$ 4.5 mag) and the development of multi-peaked H$\alpha$ emission had occurred during the period the SN was unobservable. The sharp luminosity decline between days 82 and 232  may have been caused by a low mass ($\lesssim$3 x 10$^{-3}$M$_{\sun}$) of $^{56}$Ni in the ejecta combined with the formation of new dust in the CDS.  The presence of dust is supported by the red/blue asymmetries seen in the H$\alpha$ components as well as the presence of a large IR excess. In SN 2007od, the dust formed between days 114-232 which is much earlier than seen for other Type IIP SNe, but in line with SNe that are interacting with the CSM such as SN 1998S. There is also an indication in the observed flattening of the late-time lightcurve that a light echo from surrounding circumstellar or interstellar medium is present around SN 2007od. Flux from this echo also would make a significant contribution to the H$\alpha$ emission profile at late times.

SN 2007od is unusual in that, while it is classified as a Type IIP, there is evidence for CSM lying close to the progenitor, even though it showed neither a narrow H$\alpha$ component nor X-ray emission which are both present in typical Type IIn SNe. We believe the multi-peaked H$\alpha$ emission in SN 2007od arises from a similar mechanism to that seen in SN 1998S, and is consistent with the interaction of the ejecta with a circumstellar torus and blob. Assuming that scenario, the $\pm$ 1500 km s$^{-1}$ emission arises from the ejecta interacting with a circumstellar torus, and the -5000 km s$^{-1}$ arises from a single cloud or blob of shocked CSM out of the plane of the
CSM. The use of a torus and blob is to illustrate the ways that the multicomponent H$\alpha$ emission may arise but should not be construed as the ``true'' geometry of the SN 2007od which is not well constrained by the data. 

Our radiative transfer models have indicated that up to 4.2 x 10$^{-4}$M$_{\sun}$ of dust has formed in the CDS of SN 2007od. The models indicate that 2.5 x 10$^{-4}$M$_{\sun}$ was present on day 300 and an additional 1.7 x 10$^{-4}$M$_{\sun}$ formed by day 455. This amount of new dust is similar to estimates for other CCSNe. For example SN 1987A produced 7.5 x 10$^{-4}$ M$_{\sun}$ of dust \citep{2007MNRAS.375..753E} and SN 2006jc  an estimated 3 x 10$^{-4}$M$_{\sun}$ of dust \citep{2008MNRAS.389..141M}.  Dust estimates for SN 2003gd currently yield values between  2 x 10$^{-4} $M$_{\sun}$  and 2 x 10$^{-2} $M$_{\sun}$ from \citet{2006Sci...313..196S} and 4 x 10$^{-5}$M$_{\sun}$ from \citep{2007ApJ...665..608M}, both based on the same data. There seems to be a significant variation in the amount of dust produced from SN to SN, but none of them approach the  $\sim$1 M$_{\sun}$ of dust per SN needed to account for the massive amounts of dust seen in early-time galaxies at high red-shift \citep{1989ApJ...344..325K, 2001MNRAS.325..726T}. Therefore, in the few dust-forming Type II SNe that have been observed, there is no evidence that these objects are major contributors to the dust-rich early galaxies seen at high redshift.
The sample is still very meager, and more such SNe need to
be observed with good temporal and wavelength coverage. Monitoring of SN 2007od is continuing in the optical and IR.\\
\\
We would like to thank the anonymous referees for valuable suggestions which have improved this paper. We would also like to thank St\'{e}phane Blondin, Robert Kirshner, and the rest of the CfA SN group for providing the early-time spectra of SN 2007od, and to Ryan Chornock at Berkeley for allowing us access to his early time lightcurve and spectoscopic data. This work has been supported by NSF grant AST-0707691 and NASA GSRP grant  NNX08AV36H. This work is based in part on observations made with the Spitzer Space Telescope, which is operated by the Jet Propulsion Laboratory, California Institute of Technology under a contract with NASA. Support for this work was provided by NASA through an award (RSA No. 1346842) issued by JPL/Caltech. A portion of this data was obtained at the Gemini Observatory, which is operated by the Association of Universities for Research in Astronomy (AURA) under a cooperative agreement with the NSF on behalf of the Gemini partnership. The standard data acquisition has been supported by NSF grants AST-0503871 and AST-0803158 to A. U. Landolt.

\begin{table*}[htbp]
  \centering
  \topcaption{Observation Summary} 
  \begin{tabular}{cccccc} 
  \hline
Day & JD & Telescope & Instrument  & Exposures & Time/Exposure \textit{(s)}\\
\hline
57&2454455& SMARTS 1.3m &ANDICAM& 5& 20\\
232&2454630& Gemini South& GMOS Spectra&3&900\\
232&2454630& Gemini South& GMOS Imaging&3&20\\
300&2454698&Spitzer&IRAC&12&100\\
309&2454707&Gemini South&GMOS Spectra&3&900\\
309&2454707&Gemini South&GMOS Imaging&1&60\\
348&2454746&Gemini South&GMOS Spectra&3&900\\
348&2454746&Gemini South&GMOS Imaging&2&60\\
455&2454853&Spitzer&IRS PUI&9&15\\
458&2454856&Gemini North&GMOS Spectra&3&900\\
458&2454856&Gemini North&GMOS Imaging&1&60\\
461&2454859&Spitzer&IRAC&12&100\\
471&2454869&Spitzer&MIPS&16&30\\
659&2455057&HST&ACS/WFC&4&98\\
667&2454865&Spitzer&IRAC&12&100\\
672&2455070&Gemini South&GMOS Spectra&6&1800\\
672&2455070&Gemini South&GMOS Imaging&2&60\\
699&2455097&Gemini South&GMOS Spectra&6&1800\\
709&2455107&HST&ACS/WFC&4&98\\
811&2455109&HST&ACS/WFC&4&98\\

  \hline
  \hline
  \end{tabular}
  \label{tab:booktabs}
\end{table*}

\begin{table*}[htbp]
  \centering
  \topcaption{Optical Photometry of SN 2007od} 
  \begin{tabular}{cccc} 
  \hline
Day & $V$ & $R$ & $I$\\
\hline
57& 14.86 $\pm$ 0.04 &- & 14.17 $\pm$ 0.03\\
232&21.32 $\pm$ 0.04&20.43 $\pm$ 0.02&20.21 $\pm$ 0.03\\
309&22.05 $\pm$ 0.05&21.36 $\pm$ 0.03&21.10 $\pm$ 0.05\\
348&22.44 $\pm$ 0.05&21.80 $\pm$ 0.03&21.54 $\pm$ 0.05\\
458&23.51 $\pm$ 0.12&22.66 $\pm$ 0.07 &22.77 $\pm$ 0.15\\
659&23.64 $\pm$ 0.18&- &24.08 $\pm$ 0.35\\
672&24.67 $\pm$ 0.15&23.82 $\pm$ 0.10 &24.10 $\pm$ 0.25\\
709&23.68 $\pm$ 0.21&- &24.22 $\pm$ 0.50\\
811&23.76 $\pm$ 0.20&- &24.24 $\pm$ 0.44\\

  \hline
  \hline
  \end{tabular}
  \tablecomments{ Day 57 is SMARTS/ANDICAM photometry, and days 659, 709, and 811 are HST/ACS. The remaining epochs are Gemini/GMOS photometry.}
  \label{tab:booktabs}
\end{table*}

\begin{table}[h]
\centering
\topcaption { Spitzer Photometry of SN 2007od}
\begin{tabular}{cccccccc}
\hline
Day&3.6$\mu$m&4.5$\mu$m&5.8$\mu$m&8.0$\mu$m&16$\mu$m&24$\mu$m\\
\hline
300&110.1 $\pm$ 1.6&156.4 $\pm$ 1.8&200.0 $\pm$ 7.6&199.5 $\pm$ 12.2&-&-\\
461&37.9 $\pm$ 1.4 &63.4 $\pm$ 1.6 &90.0 $\pm$ 7.5 &122.2 $\pm$ 11.8&$<$59&$<$35\\
667&23.5 $\pm$ 0.7  &33.3 $\pm$ 0.8  &-&-&-&-\\
\hline
\hline

\end{tabular}
\tablecomments{Fluxes are reported in $\mu$Jy. The 16$\mu$m photometry is from day 455 and the 24$\mu$m photometry from day 471. The day 667 IRAC photometry was obtained with warm Spitzer, which only operates in 3.6 and 4.5 $\mu$m.}
\end{table}

\begin{table}[h]
\centering
\topcaption {Blackbody Fits}
\begin{tabular}{lccccccccc}
\toprule
\multicolumn{1}{c}{} &
\multicolumn{4}{c}{Blackbody Optical} &
\multicolumn{4}{c}{Mod. Blackbody IR} \\
\cmidrule(l){2-4}
\cmidrule(l){6-8}
Epoch & T (K) & R (AU) & L (L$_{\odot}$) && T (K) & R (AU)  & L (L$_{\odot}$) && L$_{tot.}$ (L$_{\odot}$) \\
\midrule
300 d & 5100 & 5.7 & 9.2e5 & &580 & 1824 & 2.4e6 & &3.3e6 \\
455 d & 5100 & 2.9 & 2.3e5 & &490 & 1860 & 1.1e6 & &1.3e6 \\
667 d & 5100 & 1.6 & 7.5e4 & &600 & 767 & 5.1e5 & &5.8e5 \\
\bottomrule
\end{tabular}
\end{table}

\begin{table}[h]
\centering
\topcaption {Monte Carlo Radiative Transfer Models}
\begin{tabular}{lccccccccc}
\toprule
\multicolumn{2}{c}{} &
\multicolumn{6}{c}{Smooth} &
\multicolumn{1}{c}{} &
\multicolumn{1}{c}{Clumpy} \\
\cmidrule(l){3-8}
\cmidrule(r){10-10}
Epoch & AC/Si & T (K) & R$_{in}$ (AU) & R$_{out}$(AU) & L$_{tot.}$ (L$_{\odot}$) & $\tau$$_{v}$ & M$_{d}$ (M$_{\odot}$) & & M$_{d}$ (M$_{\odot}$) \\
\midrule
300 d & 0.75/0.25 & 6000 & 50.1 & 2005 & 3.4e6 & 1.50 & 1.7e-4 & & 2.5e-4 \\
455 d & 0.75/0.25 & 6000 & 50.1 & 1537 & 1.4e6 & 2.00 & 1.9e-4 & & 4.2e-4 \\
667 d & 0.75/0.25 & 6000 & 50.1 & 936& 8.1e5 & 2.75 & 1.8e-4 & &4.2e-4 \\
\bottomrule
\end{tabular}
\end{table}

 \begin{figure}[h]
\centering
   \includegraphics[width =6in]{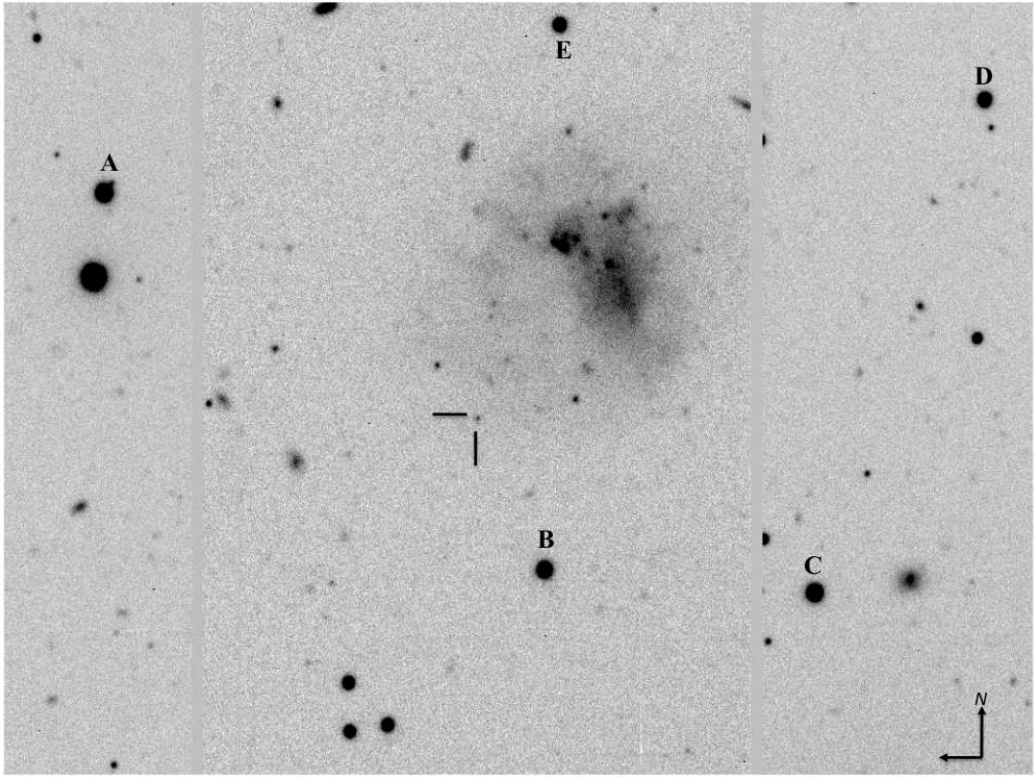}  
\caption{ The field of SN 2007od showing the location of the tertiary standards (designated alphabetically) listed in Table 6.  The location of SN 2007od is marked. North is up and east is left. The image is 4.7$^{\prime}$ x 3.1$^{\prime}$ and was taken with Gemini/GMOS-S on 2008 August 29 (day 309) with the g$^{\prime}$ filter.}
\end{figure}

\begin{figure}[htp] 
  \includegraphics [width= 2.6 in,angle=90] {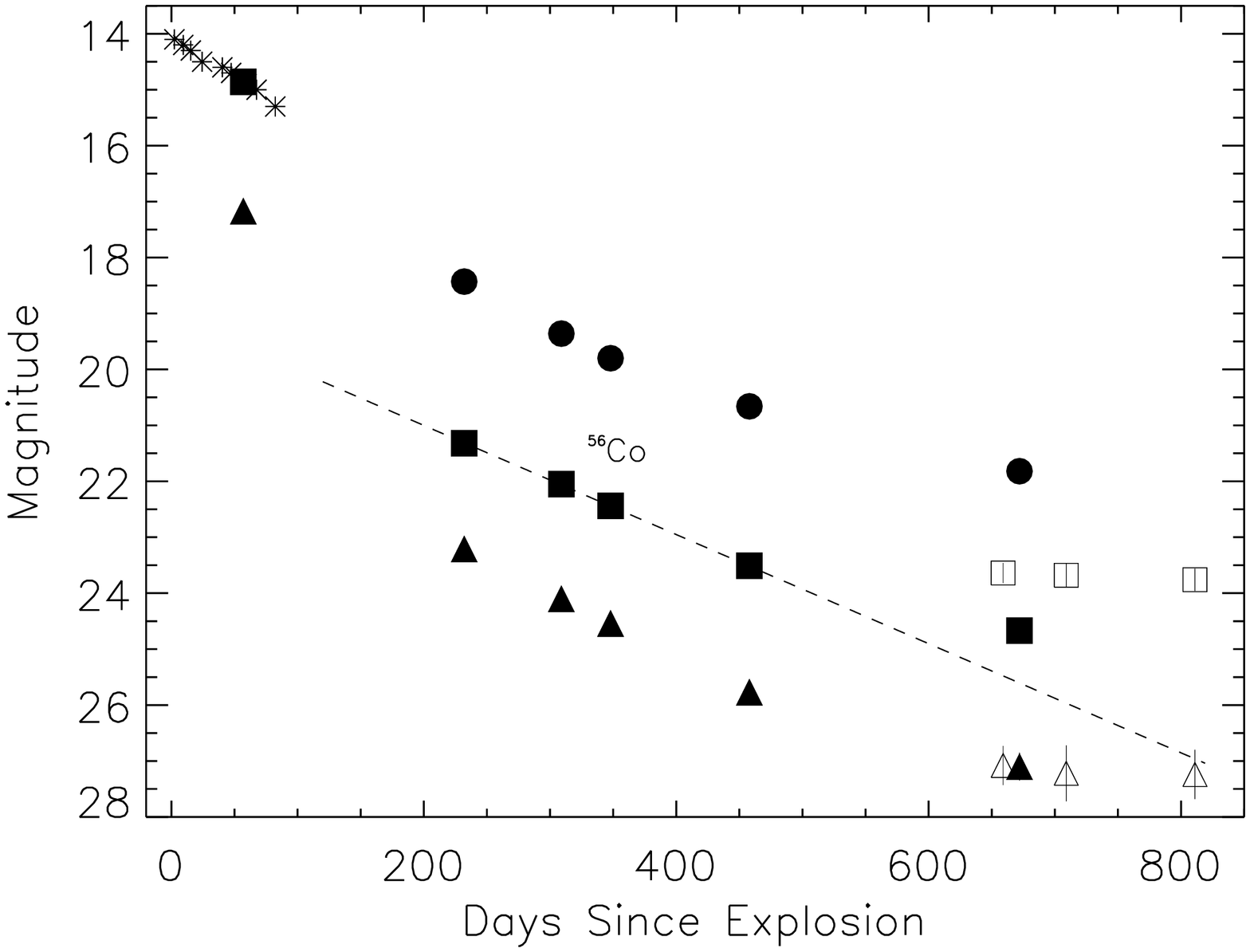} 
  \hfill
  \includegraphics  [width=2.15 in, angle=90] {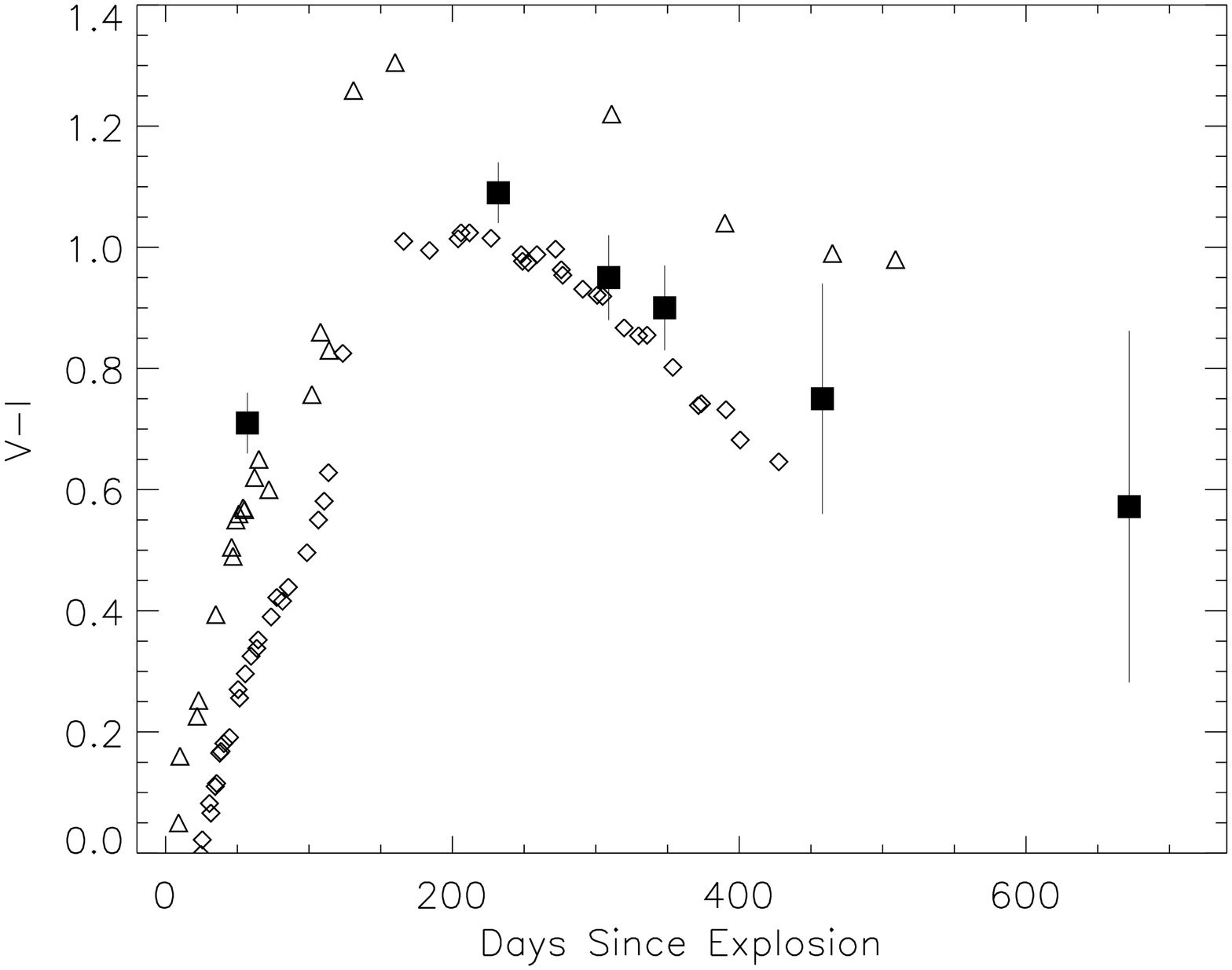} 
\caption{\textbf{Left:} Johnson-Cousins $VRI$ lightcurves of SN 2007od. $R$ points are shown as circles and have been shifted up 2 mag, $V$ are shown as squares, and $I$ as triangles shifted down 3 mag.  The filled symbols are from Gemini/GMOS and the open symbols from HST ACS/WFC. Additional photometry of SN 2007od obtained with no filter is plotted (asterisks). The uncertainties are listed in Table 2, and are in most cases smaller than symbol size. The dashed line represents the $^{56}$Co decay. Note the drop of $\sim$4.5 mag in V in 20 days, which is large for a Type IIP SN. The flattening of the lightcurve after day 600 shows the presence of a light echo which is $\sim$10 mag fainter than the SN at maximum.The disagreement in $V$ between Gemini and HST at late epochs is most likely due to filter bandpass differences (see text). \textbf{Right:} Evolution of Gemini $V-I$ of SNe 2007od (filled squares), 1999em (triangles), and 2004et (diamonds) \citep{2003MNRAS.338..939E,2006MNRAS.372.1315S}.  The $V-I$ colors of SNe 1999em and 2004et have been corrected for foreground extinction.}

 \end{figure}

\begin{figure}[h!]
\centering
   \includegraphics[width =4in, angle=90]{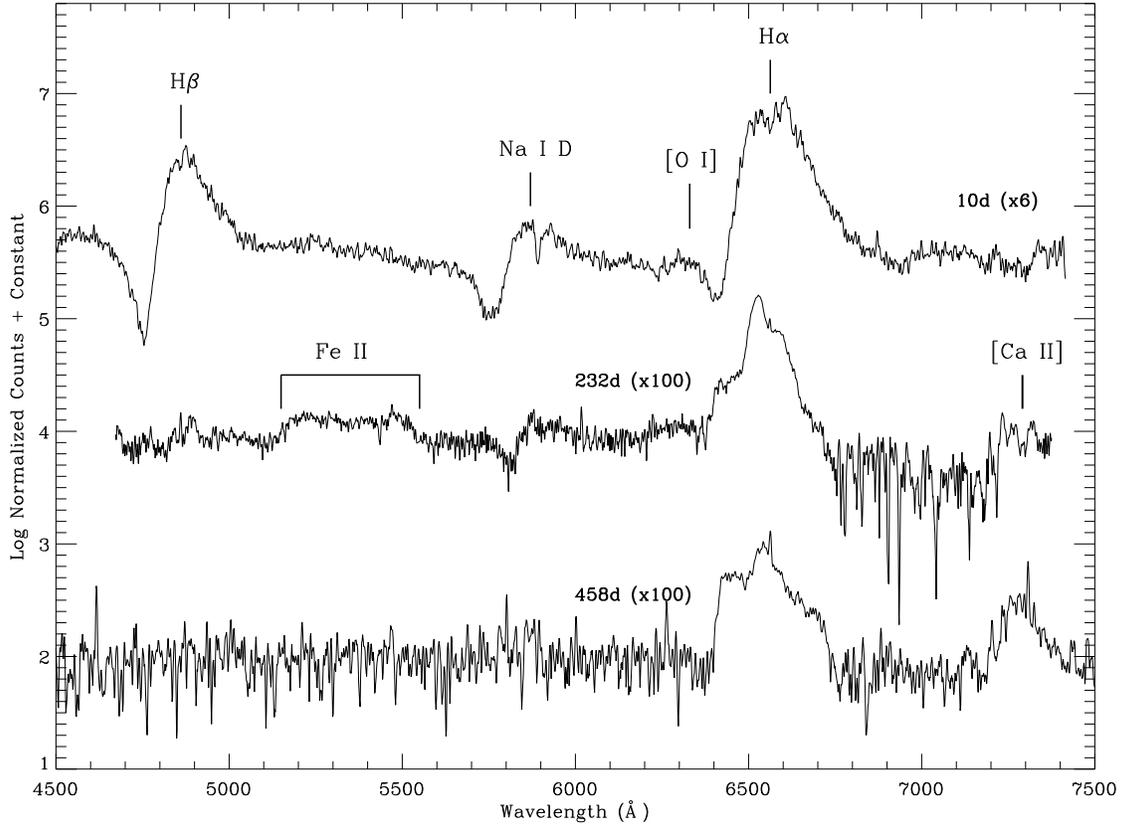}  
\caption{Optical spectra of SN 2007od on days 10, 232 and 458. All spectra have been normalized to the continuum and are shifted vertically by a constant. See text.}
\end{figure}

 \begin{figure}[h]
  \centering
    \includegraphics[height= 3.5 in]{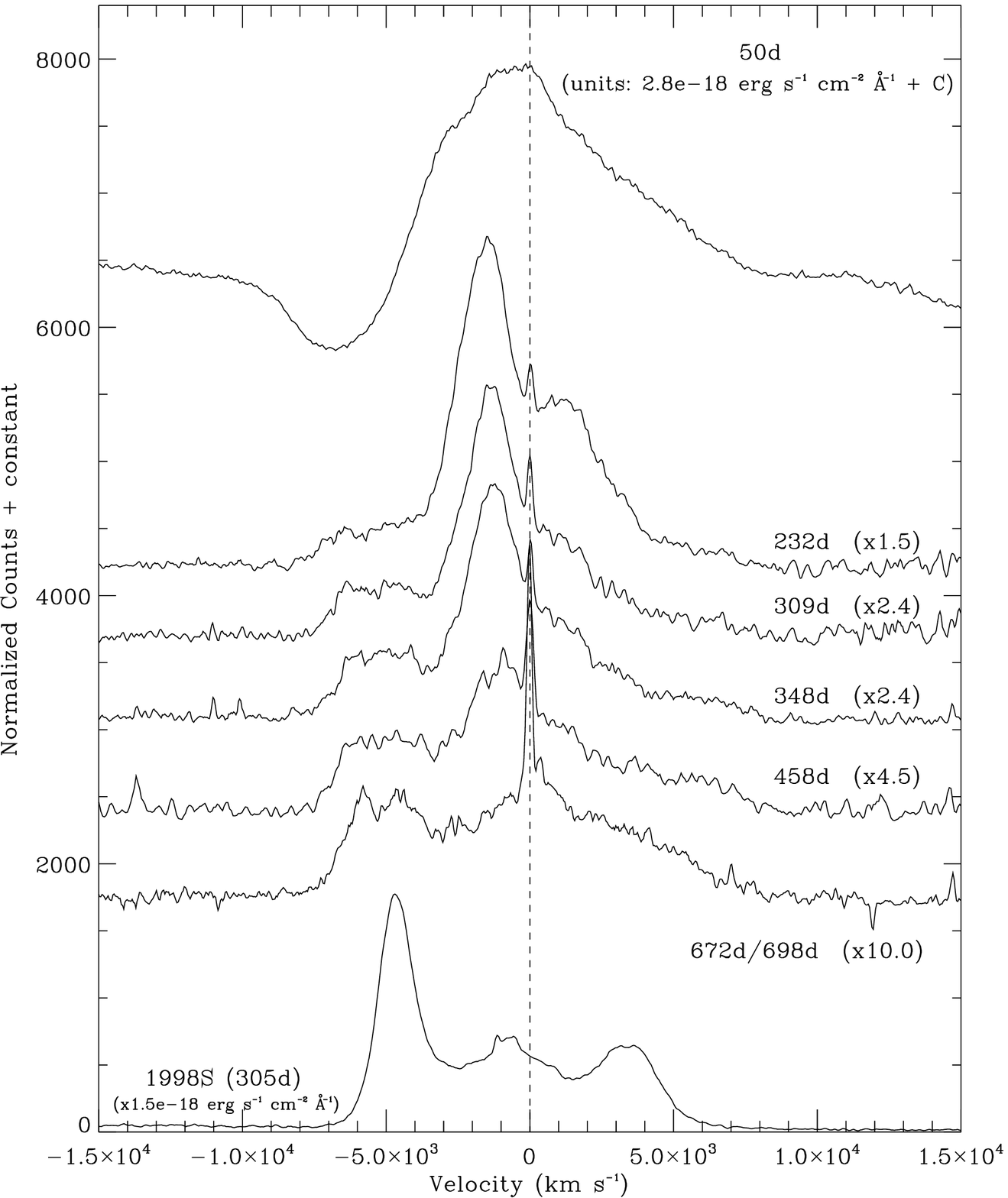}
     \includegraphics[width= 3.4 in, height=3.3 in]{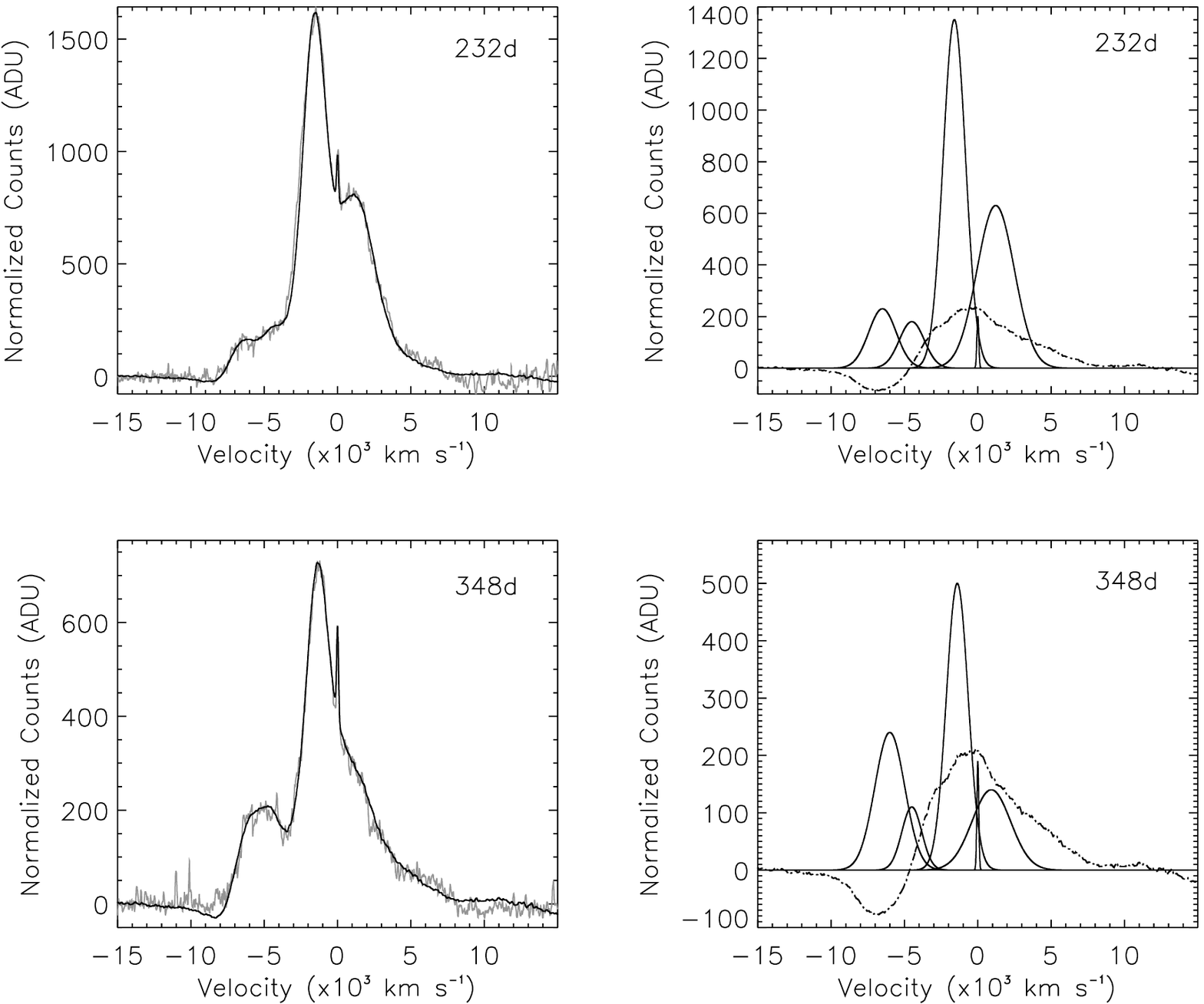} 
\caption{\textbf{Left:} Spectral evolution of the H$\alpha$ profile of SN 2007od. The -5000 km s$^{-1}$ component is easily visible by day 309, and the inner $\pm$1500 km $^{-1}$ components disappear by day 672/699. For comparison, the H$\alpha$ profile of SN 1998S (day 305) has been included \citep{2004MNRAS.352..457P}. \textbf{Right:} Deconvolution of the H$\alpha$ profiles on day 232 and 309 are shown on the right side panels, and the fit is shown as a black line over the spectrum on the left side panels.  The dashed-dotted line is a scaled day 50 profile, showing that the broad underlying emission at these epochs is consistent with being produced by directly transmitted light from the expanding ejecta.  The intermediate width $\pm$ 1500  km s$^{-1}$ and -5000  km s$^{-1}$ components are typically produced by CSM interaction as shown by comparison with the SN 1998S spectrum.  The double gaussians required to fit the -5000  km s$^{-1}$ component likely represent density enhancements in the CSM.  The day 672/699 spectra, shown on the left, is likely to be mostly due to a light echo which only became a significant contributor when the SN had faded by $\sim$ 10 mag. }
\end{figure}

\begin{figure}[h!]
\centering
  \includegraphics[width =5in]{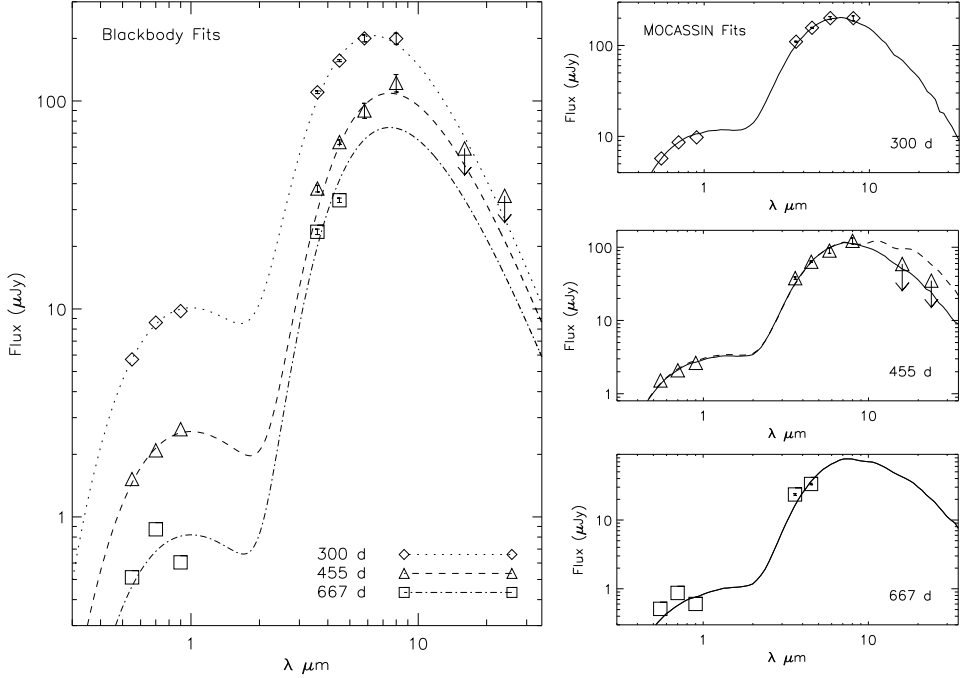}
\caption{ SED analysis of SN 2007od on days $\sim$300 (diamonds), $\sim$460 (triangles), and $\sim$667 (squares). Visible wavelengths were obtained with Gemini/GMOS and IR wavelengths with Spitzer/IRAC, IRS PUI, and MIPS.  Dust emission is clearly visible in the 3.6-8 $\mu$m wavelength range.  \textbf{Left:} SEDs with best fit blackbody curves.  The blackbody data is given in Table 4.  Optical data was fitted with a simple blackbody while the IR data was fitted with a modified blackbody subject to a $\lambda$$^{-1}$ emissivity law.  \textbf{Right:} The corresponding best fit MOCASSIN dust models for each epoch. The best fits reflected a smooth distribution of 75$\%$ AC dust (solid lines). For comparison, the best fit  75$\%$ silicate model for day 455 (dashed line) is shown and predicts too much flux past 10$\mu$m. MOCASSIN input parameters are shown in full in Table 5 for each model.}
\end{figure}

\begin{figure}[h!]
\centering
   \includegraphics[width =5 in]{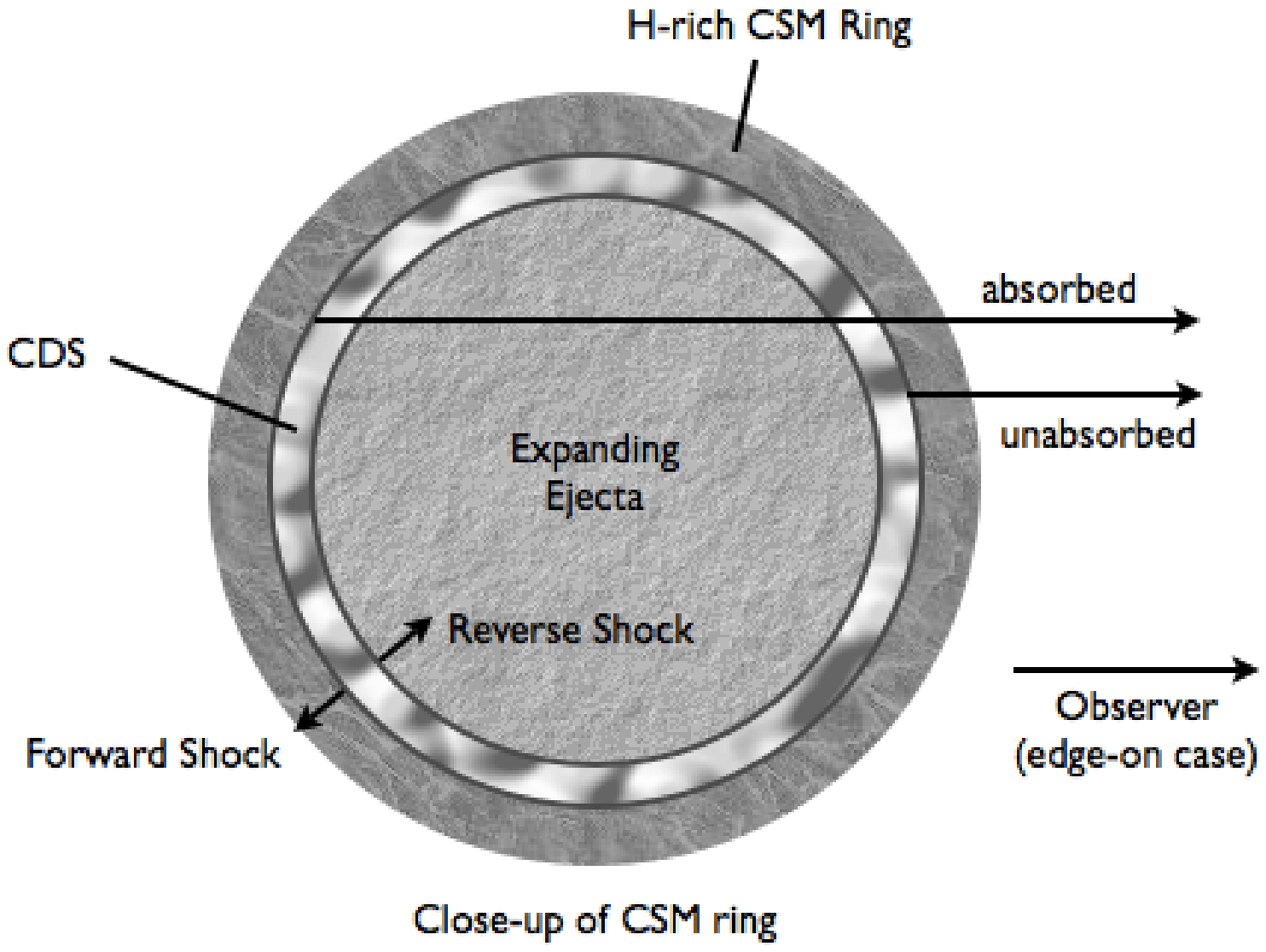}  
\caption{A cartoon showing the geometry of SN 2007od (adapted from \citet{2008MNRAS.389..141M} and \citet{2009ApJ...691..650F}).  The inner $\pm$1500 km s$^{-1}$ components may come from the forward shock created when the SN ejecta plowed into a central CSM ring, and the outer -5000 km s$^{-1}$ component from the ejecta interacting with a blob of CSM out of the plane of the central ring. If the dust is forming in the CDS between the shocks, then the emission coming from the forward shock at the CSM ring explains why the near-side emission is less reddened than the far-side emission.  See text.
} 
\end{figure}

\clearpage
\newpage
\appendix
\section{Tertiary Standards for UGC 12846}
The photometric sequence of tertiary standard stars used for both SMARTS and Gemini photometry (Table 6) was obtained using the Y4KCam CCD on the Yale 1.0m telescope operated at the Cerro Tololo Inter-American Observatory on the night of 2008 October 28.  The Y4KCam offered a field of view of $\sim$20$^{\prime}$x20$^{\prime}$ and employed a set of standard Johnson-Kron-Cousins broadband BVRI filters.  The processing of all images to remove instrumental signatures was accomplished using
the standard techniques of subtracting a median filtered bias frame and dividing by master twilight flat field images for each filter.  Also, the fringing effects in the I-band exposures were removed by scaling and subtracting a master fringe frame from the program images.  In order to transform of the instrumental magnitudes to the standard system, we observed between 30 and 35 different stars (depending on filter) from the lists of  \citet{2009AJ....137.4186L} in addition to the SN 2007od field. Importantly, these stars were selected due to their ample range in color, and they were observed both near zenith and at a high airmass ($\sim$1.75).  Thus, they provided a viable sample of stars from which to derive accurate color and extinction coeffecients as well as magnitude zero-points for each filter.  The resulting RMS scatter of the residuals from the transformations revealed that the photometry for selected stars in the SN 2007od field derived from this night's observations is accurate on the 1-2$\%$ level depending on filter.

\begin{table*}[h!]
\caption{Tertiary BVRI Standards for UGC 12846}
\centering
\begin{tabular}{ccccc}
\hline
\hline
Star&B&V&R&I\\
\hline
A&16.989 $\pm$ 0.014&15.938 $\pm$ 0.011&15.338 $\pm$ 0.006&14.820 $\pm$ 0.017\\
B&17.480 $\pm$ 0.020&16.357 $\pm$ 0.011&15.720 $\pm$ 0.006&15.153 $\pm$ 0.018\\
C&17.266 $\pm$ 0.015&16.109 $\pm$ 0.011&15.403 $\pm$ 0.006&14.786 $\pm$ 0.017\\
D&17.966 $\pm$ 0.019&17.133 $\pm$ 0.014&16.672 $\pm$ 0.010&16.343 $\pm$ 0.068\\
E&18.146 $\pm$ 0.021&17.240 $\pm$ 0.014&16.708 $\pm$ 0.010&16.267 $\pm$ 0.039\\
\hline
\end{tabular}
\centering
\end{table*}

\end{document}